\documentclass[sigconf,natbib=true]{acmart}

\acmSubmissionID{42}
\setcopyright{none}
\renewcommand\footnotetextcopyrightpermission[1]{} 
\usepackage{booktabs}
\usepackage{array}
\usepackage{balance} 
\usepackage{lipsum}
\usepackage{multirow}
\usepackage{booktabs}
\usepackage{amsmath}
\usepackage{amsthm}
\usepackage{bm}
\usepackage{enumitem}
\usepackage{setspace}
\newcommand\blfootnote[1]{%
\begingroup
\renewcommand\thefootnote{}\footnote{#1}%
\addtocounter{footnote}{-1}%
\endgroup
}

\AtBeginDocument{%
  \providecommand\BibTeX{{%
    \normalfont B\kern-0.5em{\scshape i\kern-0.25em b}\kern-0.8em\TeX}}}

\setcopyright{acmcopyright}
\copyrightyear{2018}
\acmYear{2018}
\acmDOI{XXXXXXX.XXXXXXX}

\acmConference[Conference acronym 'XX]{SIGIR'22}{June 03--05,
  2018}{Woodstock, NY}

\acmPrice{15.00}
\acmISBN{978-1-4503-XXXX-X/18/06}

\begin{document}

\title{Deep Deconfounded Content-based UGC Tag Recommendation with Causal Intervention}

\author{Yaochen Zhu$^{*}$, Xubin Ren$^{*}$, Jing Yi, Zhenzhong Chen$^{\dagger}$}
\affiliation{%
  \institution{Wuhan University, Wuhan, China}
  \country{}
}
\email{{yaochenzhu, rxubin, yijing-v, zzchen}@whu.edu.cn}

\renewcommand{\shortauthors}{Xubin Ren*, Yaochen Zhu*, Jing Yi, and Zhenzhong Chen.}

\begin{abstract}
Traditional content-based tag recommender systems directly learn the association between user-generated content (UGC) and tags based on collected UGC-tag pairs. However, since a UGC uploader simultaneously creates the UGC and selects the corresponding tags, her personal preference inevitably biases the tag selections, which prevents these recommenders from learning the causal influence of UGCs' content features on tags. In this paper, we propose a deep deconfounded content-based tag recommender system, namely, DecTag, to address the above issues. We first establish a causal graph to represent the relations among uploader, UGC, and tag, where the uploaders are explicitly identified as confounders that spuriously correlate UGC and tag selections. Specifically, to eliminate the confounding bias, causal intervention is conducted on the UGC node in the graph via backdoor adjustment, where uploaders' influence on tags leaked through backdoor paths in the graph can be eliminated for causal effect estimation. Observing that solving the adjustment with do-calculus requires integrating the entire uploader space, which is infeasible, we design a novel Monte Carlo (MC) estimator with bootstrap, which can achieve asymptotic unbiasedness provided that uploaders for the collected UGCs are \textit{i.i.d.} samples from the population. In addition, the MC estimator has the intuition of substituting the biased uploaders with a hypothetical random uploader from the population in the training phase, where deconfounding can be achieved in an interpretable manner. Finally, we establish a YT-8M-Causal dataset based on the widely used YouTube-8M dataset with causal intervention and propose an evaluation strategy accordingly to unbiasedly evaluate causal tag recommenders. Extensive experiments show that DecTag is more robust to confounding bias than state-of-the-art causal recommenders. Codes and datasets are available at \url{https://github.com/Re-bin/DecTag}.
\end{abstract}



\maketitle

\begin{small}
\begin{spacing}{1}
\blfootnote{$^*$Xubin Ren and Yaochen Zhu are co-first authors.}
\end{spacing}
\end{small}

\section{Introduction}

Recent years have witnessed the transformation of user-generated content (UGC) to people's information-sharing strategy and entertainment life \cite{cha2007tube,xie2020multimodal}. Close on the heels of UGCs is the trend of attaching tags to uploaded UGCs, which are short terms that concisely summarize the information-rich UGC features \cite{sigurbjornsson2008flickr, kim2011collaborative}. Tags serve both as short indices of the UGC and eye-catchers, which help both the uploaders to gain more exposure to the target audience and the users to more conveniently locate interesting content from numerous candidates. Since tags can be words, word combinations, or even Emojis, the candidate pool for tags is so large that it becomes difficult to select suitable tags for the generated content. Consequently, tag recommender systems, which aim to automatically recommend suitable tags to users when they upload UGCs in the platform, have become an important topic that has attracted lots of attention among researchers \cite{belem2017survey}. 

Due to the large candidate tag spaces and sparse tags associated with UGCs, collaborative filtering (CF), which recommends suitable tags for one UGC with pre-existing tags based on other UGCs with similar tags, is restricted in applications due to lack of sufficient collaborative information. Content filtering, which assigns tags for UGCs based on their content features, has become the mainstream strategy for tag recommendations \cite{belem2017survey}. Traditional content-based tag recommender systems consist of two steps \cite{wu2016tag2word,tang2019integral,maity2019deeptagrec}. First, existing UGC-tag pairs are collected from the platform to form a training database. Then, a naive matching model is designed to learn the mapping from UGC features to associated tags based on the collected UGC-tag pairs. If we define the most suitable tags based solely on the content of a UGC as the causal effects of the UGC on tags, the goal of the tag recommender should be to capture such causal effects based on collected UGC-tag pairs. However, the mapping \textit{de facto} captured by naive models is unavoidably the association between existing UGC and tags. Since uploaders simultaneously assume the role of UGC-creator and tag-selector, they are confounders where their preference for tag selections, \textit{i.e.,} the selection bias, is mistakenly captured as the causal effects. This is problematic, because the bias may cause naive models to miss important, relevant tags for UGCs neglected by uploaders, which can cost them to lose potential audiences and associated economic benefits. Therefore, it is imperative to introduce causality \cite{pearl2016causal}, which aims to eliminate confounding bias and faithfully recover causal relations among variables, into tag recommender systems.

Causality has been well-explored in item recommender systems where the targets are items such as movies, merchandise, and articles \cite{bonner2018causal,zhu2022mutually,zhu2022variational}. However, causal inference for tag recommendations, although crucial, remains under-explored among researchers. Generally, causal inference for tag recommendations faces three new challenges. (1) To eliminate uploader-induced bias, the causal intervention should be conducted upon UGCs such that only the influence of UGC content on tags, \textit{i.e.,} the causal effect, is captured by the model. However, such causal effects are \textit{per se} difficult to learn since most UGCs have no tags when recommendations are required, which renders collaborative filtering, the most effective method in item recommendations, incompetent. (2) The recommendation target, \textit{i.e.,} tags, contain richer semantic information compared to items, which leads to more complex content-matching relations between UGC and tags. (3) Most importantly, confounders in tag recommendations are UGC uploaders, which are more complicated than confounders in item recommendations, which are generally assumed to be item popularity (scalar) \cite{ZhangF0WSL021}, item attributes (categorical) \cite{ma2021multi,zhu2022deep}, user preference (categorical) \cite{wang2021deconfounded}, or historical user interactions (integral vectors) \cite{xu2021causal}. Finally, a common problem for all causal recommenders is the inaccessibility of real-world datasets due to lack of bias-free groundtruth \cite{zou2020counterfactual}. This is especially evident in tag recommendations because acquiring tags devoid of uploaders' bias is anything but realistic. Therefore, it seems impossible to evaluate the effectiveness of causal tag recommenders on large-scale real-world datasets. Tackling both the model and data side challenges for causal tag recommenders to eliminate uploaders' confounding bias becomes the main focus of this paper.

To address the above challenges, we propose a deep deconfounded content-based UGC tag recommender system, namely DecTag, based on causal intervention. We first establish a causal graph representing the relations among uploaders, UGC, and tags, where uploaders are clearly identified as confounders. To eliminate the confounding bias, causal intervention is conducted upon UGC from the causal graph via backdoor adjustment with do-calculus. Previous do-calculus solvers can only guarantee unbiasedness if tags are linearly dependent upon uploaders. However, complex relations exist among UGCs, uploaders, and tags, resulting in a large extra bias in tag predictions for these methods. Therefore, we propose a novel Monte Carlo (MC)-based do-calculus approximator with bootstrap, which can achieve asymptotic unbiasedness provided that the uploaders in the collected dataset are \textit{i.i.d.} samples from the population. In addition, the proposed MC-based solver has the intuition of replacing the biased uploader with a hypothetical random uploader in the model training phase, which allows deconfounding to be achieved in an explainable manner. Finally, DecTag is model-agnostic and can be plugged into any base tag recommender system, which shows its broad potential applications. Another main contribution of this paper is that we establish a large-scale real-world tag recommendation dataset, \textit{i.e.,} YT-8M-Causal, based on the widely used YouTube-8M dataset to evaluate causal tag recommenders. Specifically, we conduct causal interventions on the test set based on substitute confounders, where we demonstrate that if the attributes that we intervened upon have equivalent influences to the confounders, it can indirectly demonstrate whether confounding bias is eliminated without the requirement of bias-free tags. The main contribution of this paper can be summarized as:

\begin{itemize}[leftmargin=2em]
    \item We introduce causality to address uploader-induced confounding bias in tag recommendations. Different from traditional methods, we establish a structural causal graph to depict the relations among uploaders, UGCs, and tags, where uploaders can be identified as confounders that spuriously correlate UGCs and tags. To the best of our knowledge, this is the first work to investigate tag recommendations from a causal perspective.
    \item Confounding bias induced by UGC uploaders is tackled by conducting causal intervention on UGC in the causal graph with backdoor adjustment. The intractable integration over the entire infinite uploader space with do-calculus is addressed by a novel Monte-Carlo estimator with bootstrap, which is guaranteed to be asymptotically consistent and unbiased provided that uploaders are random \textit{i.i.d.} samples from the population. 
    \item The proposed MC-based do-calculus approximator has an intuitive explanation that uses a hypothetical random uploader to replace the true uploader with her own selection bias when training the tag recommender. This allows a transparent deconfounding procedure with good interpretability.
    \item  A tag recommendation dataset with causally intervened test sets, which is named as YT-8M-Causal, is established based on the widely used YouTube-8M dataset, with evaluation procedures introduced accordingly to evaluate causal tag recommender systems without the requirement of unbiased tags. Experiments demonstrate that DecTag is more robust to confounding bias than state-of-the-art deconfounding strategies.
\end{itemize}

\section{Causal View of Tag Recommenders}

In this section, we formally define the task of tag recommendation from a causal perspective. Specifically, a causal graph is established where uploaders are clearly identified as confounders that spuriously correlate UGC features and tags. In addition, we introduce backdoor adjustment to block uploaders' influence such that the causal effects of UGC content on tags can be properly estimated.

\subsection{Problem Formulation}

In this paper, we mainly investigate uploader-UGC-tag triplets in the form of $\mathcal{D} = \{(\bm{u}, \bm{c}, \bm{t})_{i}, i \in \mathrm{range}(N)\}$\footnote{In this paper, italic bold symbols $\bm{u}, \bm{c} ,\bm{t}$ represent vectors, calligraphic symbols $\mathcal{U}, \mathcal{C} ,\mathcal{T}$ represent sets, and capital non-boldface symbols $U$, $C$, $T$ represent  random vectors, with exceptions such as $N$ and $N^{s}$ easily identifiable from the context.}. The triplets are collected from online UGC sharing platforms, where $\bm{u} \in \mathcal{U}$ is the representation of a selected uploader from the population, $\bm{c}$ is the feature of a content generated by uploader $\bm{u}$, and $\bm{t}$ is one of the tags that uploader $\bm{u}$ selects for the content. Since the uploader $\bm{u}$ simultaneously creates the UGC content and selects its tags, her personal bias unavoidably exists in the collected triplets, which should be eliminated when estimating UGC content's influence on tags for unbiased content-based tag recommendations. To achieve such a purpose, $\mathcal{D}$ needs to include multiple UGCs from each uploader such that the bias can be properly modeled and eliminated. Mathematically, $\bm{D}$ need to have the property that in $(\bm{u}, \bm{c}, \bm{t})_{i}$ there exists $\bm{u}_{j} = \bm{u}_{k}$ such that $\bm{c}_{j}  \neq \bm{c}_{k}, \ \ \forall \bm{u}_{j} \in \mathcal{U}$. The goal of this paper is to recommend suitable tags for newly uploaded UGCs based solely on their content features, irrespective of any specific uploader's personal preference, such that the pervasive confounding bias for tag selection in the collected datasets can be eliminated.

\subsection{Causal Graph Modeling}
\label{sec:causal_graph}

\begin{figure}
\setlength{\abovecaptionskip}{0.3cm}
\setlength{\belowcaptionskip}{-0.4cm}
\centering
\includegraphics[width=0.96\linewidth]{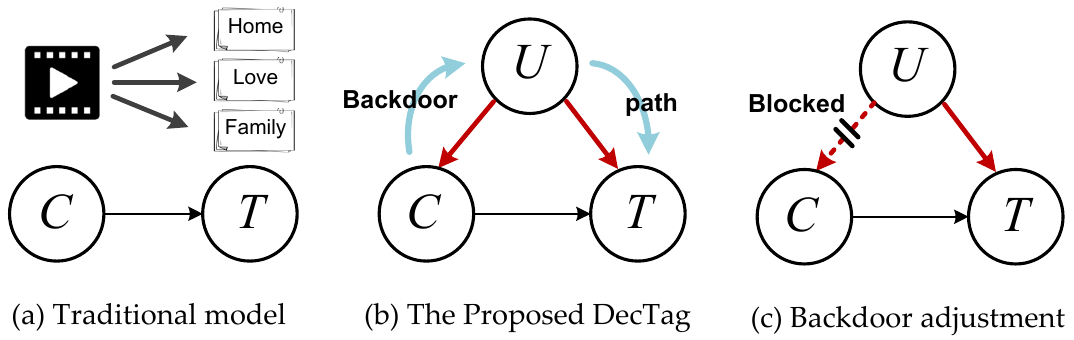}
\caption{Comparisons between causal graphs assumed by traditional tag recommenders, by the proposed DecTag, and the causal graph of DecTag after backdoor adjustment.}
\label{fig:causal_graph}
\end{figure}

To gain more intuition of the problem, we represent the causal relations among $U$, $C$, $T$ as a graph in Fig. \ref{fig:causal_graph}b, where links demonstrate the direction of causal influence between a node pair. In Fig. \ref{fig:causal_graph}b, link $U \rightarrow C$ depicts the influence of the uploader's specialized area, personal background, \textit{etc.,} on the content of the uploaded UGC, link $U \rightarrow T$ depicts the influences of the uploader's personal background, social status, characteristics, and preference, \textit{etc.,} on her selected tags, and link $C \rightarrow T$ depicts the influence of UGC's content features to tag selections. In this paper, the influence represented by link $C \rightarrow T$ is defined as the causal influence of UGC's content features on tags, of which we aim to recover from the collected U-C-T triplets, such that bias-free tag can be recommended for newly uploaded UGCs based solely on their content. 

Traditional tag recommender systems assume the simple relationship between $C$, $T$ represented by Fig. \ref{fig:causal_graph}a, which neglects the influence of UGC uploaders as confounders. Consequently, when such models are fit on a UGC-tag dataset generated according to Fig. \ref{fig:causal_graph}b, the uploader's selection bias is mistakenly captured as the causal effect of $C$ on $T$. Consider the following toy example of uploader-induced confounding bias. Suppose that Alice is an outdoor streamer who constantly uploads videos about her outdoor activities. As a naturalist, she prefers to select tags that describe the beautiful natural view of the places where she shoots the videos instead of tags that describe the sports she chooses to stream about. If a tag recommendation model trained on Alice's data unintentionally captures the spurious association between outdoor videos and tags regarding natural views as the causal effect of the video content on the tags, when Bob, who is also an outdoor streamer, uploads a video regarding outdoor activities (such as parkour or skateboard) on a busy street at downtown, Alice's bias will lead the system to recommend nature-associated tags for Bob's content, which will cause Bob to lose potential audiences who use sport-related tags to index the outdoor streaming content that suits their interests. Theoretically, such an example vividly illustrates the confounding bias associated with an open backdoor path where the influence of C can be leaked through $C \leftarrow U \rightarrow T$ on $T$. Unblocked backdoor paths can lead to systematic bias in estimated causal effects for traditional non-causality-based tag recommendation models. 
 
\section{Methodology}

We have clearly demonstrated how the uploader's personal bias can create a spurious correlation between UGC content and tags via backdoor path in the previous section. In this section, we introduce the proposed deep deconfounded tag recommender, DecTag, for unbiased content-based tag recommendations for UGCs.

\subsection{Deconfound via Causal Intervention}

To estimate the causal effects of UGC's content $C$ on tag $T$, Pearl's causality theory \cite{pearl2016causal} requires us to control the confounders such that backdoor paths are blocked. This is achieved through backdoor adjustment on UGC node $C$ of the causal graph in Fig. \ref{fig:causal_graph}b. {According to the theory of causal inference \cite{pearl2016causal},} the adjustment is formulated via do-operator as $P(T \mid do(C=\bm{c}))$, which can be intuitively understood as cutting off the edge $U \rightarrow C$ in the causal graph such that the effects of $U$ on $T$ leaked through path $C \leftarrow U \rightarrow T$ can be blocked for causal effect estimation (see Fig. \ref{fig:causal_graph}c). The specific form of backdoor adjustment is formulated as:
\begin{subequations}
\label{eq:backdoor_adj}
\begin{align}\footnotesize
&P(T \mid do(C=\bm{c})) \notag \\
                &= \int_{\mathcal{U}} P(\bm{u} \mid do(C=\bm{c}))P(T \mid do(C=\bm{c}),\bm{u})  d \bm{u}\\
                &= \int_{\mathcal{U}} P(\bm{u})P(T \mid \bm{c},\bm{u})  d \bm{u},
\end{align}
\end{subequations}
\noindent where Eq. (\ref{eq:backdoor_adj}a) is based on the law of total probability and the Bayesian rule, and Eq. (\ref{eq:backdoor_adj}b) follows the definition of do-operator \cite{pearl2016causal} and conditional independence relationships induced by the causal graph. In contrast, traditional content-based tag recommender systems fits models according to conditional distribution of $P(T \mid C=\bm{c})$ (\textit{i.e.,} causal graph Fig \ref{fig:causal_graph}a) on data actually generated according to Fig \ref{fig:causal_graph}b, which can be formulated as
\begin{subequations}
\label{eq:conv}
\begin{align}\footnotesize
&P(T \mid C=\bm{c}) = \int_{\mathcal{U}} P(\bm{u} \mid \bm{c})P(T \mid \bm{c},\bm{u})  d \bm{u}.
\end{align}
\end{subequations}
By comparing Eq. (\ref{eq:backdoor_adj}) and Eq.(\ref{eq:conv}) we can find that through Eq. (\ref{eq:backdoor_adj}), DecTag considers the tag $\bm{t}$ for content $\bm{c}$ the with every possible uploader $u$ instead of the true uploader draw from $P(\bm{u} \mid \bm{c})$. Therefore, the bias incurred by one specific uploader can be eliminated.

\subsection{Monte Carlo Sampler with Bootstrap}

Despite the guarantee to eliminate the uploader-induced confounding bias, solving Eq. (\ref{eq:backdoor_adj}) is intractable because of the integration over the infinite uploader space, which precludes us from calculating an analytical solution. Therefore, an approximation strategy is introduced to form estimators for the do-calculus integral.

\subsubsection{Approximation with Hypothetical Averaged Uploader} If we define the expectation of uploader representation $\bm{u}$, \textit{i.e.,} $ \int_{\mathcal{U}} P(\bm{u})  d \bm{u}$ as $\hat{\bm{u}}$, the estimator proposed by DecRS \cite{wang2021deconfounded} can be formulated as
\begin{equation}
\int_{\mathcal{U}} P(\bm{u})P(T \mid \bm{c},\bm{u})  d \bm{u} \approx P(T \mid \bm{c},\hat{\bm{u}}).
\end{equation}
\noindent Theoretical details of the approximation can be referred to \cite{wang2021deconfounded}. Here, we provide an intuitive explanation for the deconfounding mechanism of DecRS. We observe that substituting $\bm{u}$ with $\hat{\bm{u}}$ in conditional distribution $P(T \mid C,U)$ can be viewed as using a \textit{hypothetical averaged uploader} $\hat{\bm{u}}$ as the surrogate for the true uploader $\bm{u}$ with selection bias when training the model. Therefore, if $\hat{\bm{u}}$ indeed possesses less bias than an individual uploader $\bm{u}$, the influence of uploaders as confounders can be diminished by the substitution. 

However, one problem of this approximation strategy is that it is theoretically unclear whether the hypothetical averaged uploader $\hat{\bm{u}}$ is indeed unbiased. Consider the toy example that we have discussed in Section \ref{sec:causal_graph}. Assume that one group of outdoor streamers is biased towards selecting nature-related tags for their videos, whereas another group of streamers is biased towards using sports-related tags. If the size of the first group is larger than that of the second group, it can be expected that the hypothetical averaged uploader $\hat{\bm{u}}$ is also biased towards using nature-related tags. We call this bias the residual bias in the averaged uploader. This bias cannot be addressed by \cite{wang2021deconfounded}. Furthermore, if the conditional distribution of tags $P(T \mid C,U)$ is highly non-linear for the uploader $U$, which is common in tag recommendation due to the complex relations among uploaders, UGCs, and tags, the approximation could be biased even if the biases of all uploaders can be negated by averaging. In summary, we can determine that when the conditional distribution $P(T \mid C,U)$ is non-linear in $U$, the estimator of DecRS is neither asymptotically consistent nor unbiased for the do-calculus in Eq. (\ref{eq:backdoor_adj}).

\subsubsection{Approximation with Hypothetical Random Uploader} To address the problems, we design a Monte-Carlo (MC)-based approximator with bootstrap in this section. We note that the integral in Eq. (\ref{eq:backdoor_adj}b) can be rewritten into an expectation form as follows,
\begin{equation}
\label{eq:monte}
\int_{\mathcal{U}} P(\bm{u})P(T \mid \bm{c},\bm{u}) d \bm{u} = \mathbb{E}_{P(U)}[P(T \mid \bm{c},U)].
\end{equation}
\noindent With $P(T \mid \bm{c},\bm{u})$ explicitly formulated in an expectation form, Monte Carlo method can be introduced to form unbiased estimator for the LHS of Eq. (\ref{eq:monte}). Specifically, we first draw $N^{s}$ uploaders from $P(U)$ as $\bm{u}_{i} \sim P(U), i \in \mathrm{range}(N^s)$, where $N^{s}$ is the sample size. Then, the expectation in Eq. (\ref{eq:monte}) can be unbiasedly estimated by
\begin{equation}
\mathbb{E}_{P(U)}[P(T \mid \bm{c},U)] \approx  \frac{1}{N^{s}}\sum_{i=1}^{N^{s}} P(T \mid \bm{c},\bm{u}_{i}),
\end{equation}
where the increase of $N^{s}$ linearly reduces the estimation variance while increasing the computational complexity simultaneously. The deconfounding mechanism of the proposed DecTag has an intuition of using a \textit{hypothetical random uploader} with probability distribution P(U) to substitute the biased real uploader $\bm{u}$, where the expectation of tags provided by the random uploader are calculated as the predictions. Compared to DecRS, the randomness in DecTag could guarantee the elimination of the uploader-induced confounding bias even if $P(T \mid C,U)$ is highly non-linear for uploader $U$. 

\begin{figure}
    \centering
    \includegraphics[width=0.99\linewidth]{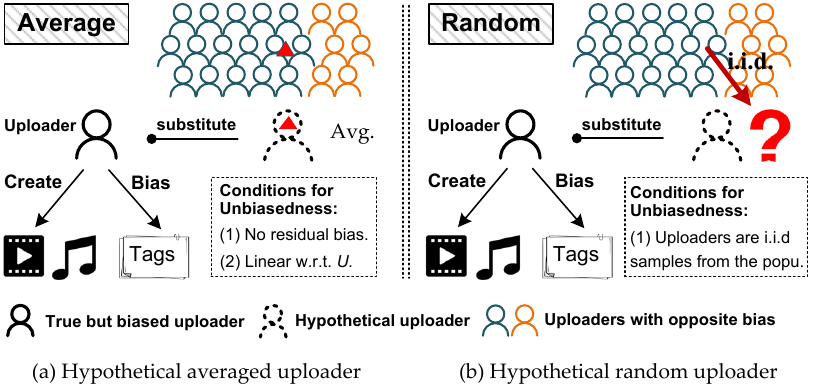}
    \caption{Comparisons between hypothetical averaged uploader used in DecRS and hypothetical random uploader proposed by DecTag for deconfounding uploaders' bias.}
    \label{fig:decrs_vs_dectag}
\vspace{-2mm}
\end{figure}

Since the uploader distribution $P(U)$ is complex, it cannot be confidently estimated from the dataset. Therefore, another problem remaining to be addressed is how to draw multiple samples from the true uploader distribution $P(U)$. In this paper, we resort to the bootstrap strategy \cite{rubin1981bayesian}. Specifically, we draw surrogate samples from the empirical uploader distribution, which is defined as $\hat{P}(U=\bm{u}) = \sum_{i=1}^{N} \mathbb{I}(\bm{u}_{i} = \bm{u}) / N$, with replacement. $\mathbb{I}$ is the identity function. The MC-estimator with bootstrap can achieve asymptotic consistency and unbiasedness provided that the uploaders in the collected datasets are \textit{i.i.d.} samples from the population \cite{rubin1981bayesian}. This can be easily guaranteed in data collection, where we first randomly sample uploaders as seed uploaders and from each seed uploader randomly selected uploaded UGCs from their collections. The comparison between the deconfounding strategy of DecRS and the proposed DecTag is intuitively illustrated in Fig. \ref{fig:decrs_vs_dectag} for reference.

\subsubsection{Bias-Variance Trade-off between DecRS and DecTag}
\label{sec:trade-off}

Please note that the asymptotically consistency and unbiasedness properties of DecTag are never a free lunch. The estimator of DecRS has a low variance by adopting a deterministic hypothetical averaged uploader $\hat{\bm{u}}$ as the surrogate for the true uploader. This may have an advantage if the datasets are small-scale or extremely sparse. In contrast, the stochastic MC-based sampling strategy introduced in DecTag requires a larger sample size $N^{s}$ to obtain equally low variance in estimation. Therefore, a higher training and testing complexity is generally expected for DecTag compared with DecRS. Previous work shows that it is possible to set $N^{s}=1$ to achieve a low variance provided that the batch size of uploaders is large \cite{kingma2013auto}. This strategy is adopted for training DecTag due to its simplicity. 

\subsubsection{Specific Form of $\bm{u}$ as Confounders}
\label{sec:conf_form}
Since only conditional distributions can be faithfully estimated from confounded datasets \cite{pearl2016causal}, we first fit $P(T \mid C,U)$ on dataset $\mathcal{D}$ for $N_{warm}$ epochs to warm up the model. After the burn-in phase, the hypothetical random uploader of the MC-based estimator is then used as the surrogate for true biased uploaders in $P(T \mid C,U)$ as Eq. (\ref{eq:monte}) to eliminate the uploader-induced confounding bias in tag selections.

In this paper, we assume that the collected UGCs are \textit{i.i.d.} samples from the collection of each uploader. This assumption implies that the sampling UGC distribution can faithfully represent the true UGC distribution of the uploaders. Based on this, we calculate the statistics of the collected UGCs of an uploader as her representation $\bm{u}$. Specifically, we first obtain the topics of each UGC included in the dataset (This can be easily acquired since most platforms, such as YouTube and Instagram, \textit{etc.,} have categorized contents based on topics). We then randomly initialize trainable UGC topic embeddings $\mathbf{W}_{t} \in \mathbb{R}^{K \times N_{tp}}$ and calculate the ratio of each topic in uploaders' collections, where $N_{tp}$ is the number of UGC topic categories and $K$ is the dimension of topic embeddings. If we define $Tp \in \mathrm{range}(1, N^{t})$ as the $Tp$th topic category, $\mathcal{C}_{\bm{u}}$ as the set of collected UGCs for uploader $\bm{u}$, and $Cat(\bm{c})$ as the function that returns the topic category of UGC $\bm{c}$, the $Tp$th dimension of $\bm{u}$'s UGC topic distribution $\hat{\bm{u}}$ can be calculated by $\hat{\bm{u}}_{Tp} = \sum_{\bm{c} \in \mathcal{C}_{u}} \mathbb{I}(Cat(\bm{c})=Tp) \ / \ |\mathcal{C}_{u}|$. The uploader representation is then calculated as $\bm{u} = \mathbf{W}_{t} \cdot \hat{\bm{u}}$, which is the weighted sum of UGC topic embeddings. Deep latent variable model can be further introduced by viewing $\bm{u}$ as noisy observations of latent confounders $\bm{z}_{u}$ when $\bm{u}$ is a sparse vector lies in a high-dimensional space \cite{louizos2017causal}. However, this is beyond the scope of this paper and can be explored as future research.

\subsection{Model-Agnostic Training Strategy}

The derivation of the deconfounding ability of DecTag in Eq. (\ref{eq:monte}) is based solely on the abstract conditional distribution $P(T \mid C,U)$, which shows that DecTag is model-agnostic and can be built upon arbitrary content-based tag recommender systems. For DecTag to materialize, the functional form that parameterizes $P(T \mid C,U)$ remains to be specified. Theoretically, any function $\mathrm{f}_{\boldsymbol{\theta}}: \mathcal{C} \times \mathcal{U} \rightarrow \mathcal{T}$ with signature of $U$, $C$ as inputs, probabilities for predicted tags $T$ as outputs, and $\boldsymbol{\theta}$ as trainable parameters can be used to parameterize the distribution. In this paper, we assume that $P(T \mid C,U)$ decomposes into two compact factors as $\frac{1}{E} \hat{P}(T \mid C) \cdot \hat{P}(T \mid U)$ to explicitly disentangle uploader's bias towards tags and the influence of UGC's content to tags {\cite{cadene2019rubi, niu2021counterfactual}}. $\hat{P}$ represents unnormalized probability distribution, and $E$ is the normalization constant. 

\subsection{Base Models for Implementations}

Specifically, we adopt two widely-used strategies, \textit{i.e.,} neural factorization machine (NFM) \cite{he2017neural} and LightGCN \cite{he2020lightgcn}, to implement the two decomposed factors of $P(T \mid C,U)$, \textit{i.e.,} $\hat{P}(T \mid C)$ and $\hat{P}(T \mid U)$. {Although both methods are CF-based originally, they can be easily adapted for content-based recommendations by replacing the UGC ID embeddings with their content features}. To avoid redundant discussions, we define $X \in \{C, U\}$. In our implementation of NFM, the two unnormalized probabilities are calculated as:
\begin{equation}
\label{eq:nfm}
  \hat{P}(\bm{t} \mid \bm{x}) = f_{out}\Big( M \Big(M(\bm{t}) \odot M(\bm{x}) \ || \ M\big( M(\bm{t}) \ || \ M(\bm{x})\big)\Big) \Big),
\end{equation}
where $f_{out}$ is the identity function $f_{out}(\bm{x_{in}}) = \bm{x_{in}}$ if $X=C$ and is the sigmoid function $f_{out}(\bm{x_{in}}) = 1 / (1 + e^{-\bm{x_{in}}})$ if $X=U$, $M$ represents the multi-layer perceptron, $\bm{t}$ is the one-hot representation of the target tag, $\odot$ is element-wise product, and $||$ represents the concatenation operator. Another backbone model, \textit{i.e.,} LightGCN, is a state-of-the-art graph neural network introduced in \cite{he2020lightgcn}. In LightGCN, a bipartite graph is first established between $T$, $C$ based on the collected U-C-T triplets. For the graph, messages from UGC content features $C$ to tags $T$ are then propagated as follows:
\begin{equation}
\label{eq:gcn}
\bm{t}_{i}^{(k+1)}=\operatorname{AGG}\left(\bm{c}_{j}^{(k)},\left\{\bm{c}_{j}^{(k)}: \bm{c}_{j} \in \mathcal{N}_{\bm{t}_{i}}\right\}\right) ,
\end{equation}
where $\operatorname{AGG}$ is the LightGCN aggregation function defined in \cite{he2020lightgcn}, $i$, $j$, $k$ are used to distinguish different tags, UGC content, and representations from different GCN layers, respectively. $\mathcal{N}_{\bm{t}_{i}}$ denotes the set of neighborhood for tag $\bm{t}_{i}$. Similar message passing mechanism exists from $T$ to $C$. Finally, inner product between aggregated tag representations and UGC content representations is calculated as unnormalized probabilities for $\hat{P}(T \mid C)$. $\hat{P}(T \mid U)$ is calculated by the inner product between uploader representation $\bm{u}$ and tag embeddings from the first GCN layer learned on the $T-C$ graph.  

\subsection{Tag Prediction for Newly Uploaded UGCs}

After training, the learned $P(T \mid C,U)$ can be used for deconfounded tag recommendations. Specifically, when uploader $\bm{u}$ uploads a new UGC $\bm{c}_{new}$ to the platform, Eq. (\ref{eq:monte}) can be used to form unbiased tag predictions for $\bm{c}_{new}$. This can be intuitively understood as randomly selecting $N^{s}$ uploaders from the population and calculating averaged tag predictions. The averaged scores are then ranked where the top-$K$ tags are selected for recommendations.

\section{YT-8M-Causal Dataset}

It is notoriously difficult to evaluate deep causal recommenders on real-world datasets due to the lack of groundtruth. The problem is especially evident for the tag recommendation task considered in this paper because the bias-free tags for each UGC, (\textit{i.e.,} the tags that best fit the UGC based solely on its content) is inherently inaccessible. Despite the difficulties, we established a large-scale real-world tag recommendation dataset, namely YT-8M-causal, based on the widely adopted YouTube-8M dataset. With a dataset intervention strategy introduced accordingly, the deconfounding ability of different strategies can be indirectly evaluated and compared. 

\subsection{Data Collection Before Intervention}

\begin{table}[t]
\caption{Statistics of the established YT-8M-Causal dataset.}
\label{tab:statistic_dataset}
\resizebox{.47\textwidth}{!}{
\begin{tabular}{l|l|l|l|l|l}
\hline
\textbf{Topics} & \textbf{\#Users} & \textbf{\#Videos} & \textbf{\#Tags} & \textbf{\#Tag Usage} & \textbf{\#Feature} \\ \hline
\textbf{P.A. \& H.L.} & 17,115 & 38,442 & 6,308 & 244,340 & 1,024 \\ \hline
\textbf{A.V. \& B.I.} & 21,686 & 16,845 & 7,203 & 339,856 & 1,024 \\ \hline
\end{tabular}
}
\vspace{-0.4cm}
\label{tb:stats}
\end{table}

YouTube-8M \cite{abu2016YouTube-8M} is one of the largest multi-label video classification datasets collected from YouTube, which is a world-famous video-sharing platform with billions of active users. The dataset consists of 8 million videos from 24 categories of topics, where for each video, a 1,024-dimensional Inception-net-based feature is extracted as the video content representation $\bm{c}$. {Other features, such as context features, can also be included as \cite{tang2019integral}. However, we mainly focus on UGC contents such that the impact of uploaders' confounding bias on content-based tag selections can be better investigated.} To establish the YT-8M-Causal dataset. We first select two groups of videos in four first categories, \textit{i.e.,} Pets \& Animals (P.A.) and Hobbies \& Leisure (H.L.) for the first group, and Autos \& Vehicles (A.V.) and Business \& Industrial (B.I.) for the second group, to form two sub-sections of the dataset. We then collect the uploader-selected tags of each video via the YouTube Data API (v3). Filtering is applied on tags in each sub-dataset to ensure the quality of recommended tags, where tags with less than two letters or used by less than 20 videos are deemed as low-quality and are eliminated. After that, videos with no tags are excluded from the dataset. Finally, we remove uploaders with less than five uploaded videos to ensure the quality of uploader representations. The dataset statistics are in Table \ref{tb:stats}. 

\subsection{Analysis of the Pre-processed Dataset}

\begin{figure}
\setlength{\abovecaptionskip}{0.3cm}
\setlength{\belowcaptionskip}{-0.35cm}
\centering
\includegraphics[width=0.76\linewidth]{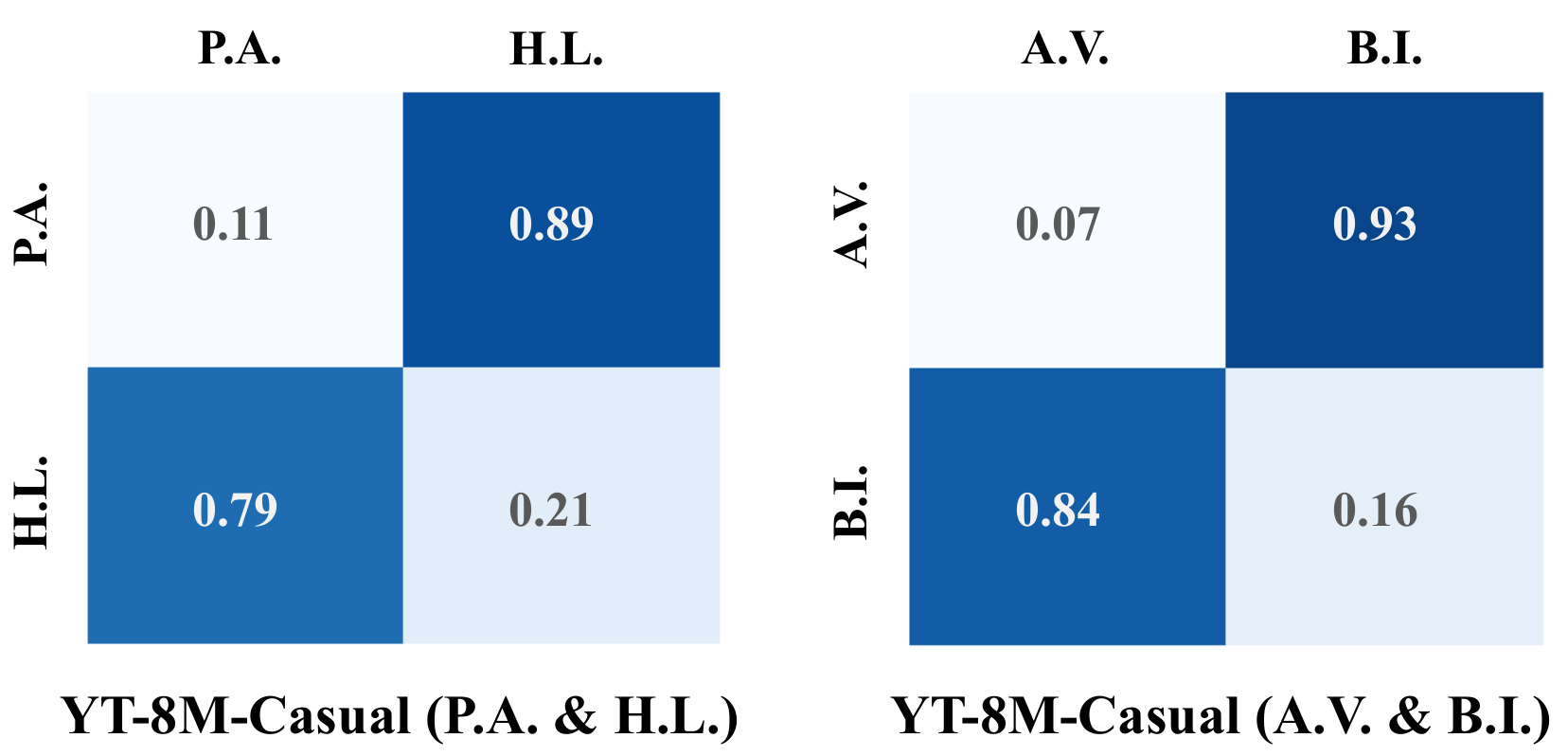}
\caption{The $ii$th item of the matrices shows the percentage of tags inclusive for topic $i$, whereas the $ij$th item shows the percentage of mutual tags for topics $i,j$.}
\label{fig:comatrix}
\end{figure}

\begin{figure}
\setlength{\abovecaptionskip}{0.2cm}
\setlength{\belowcaptionskip}{-0.2cm}
\centering
\includegraphics[width=0.9\linewidth]{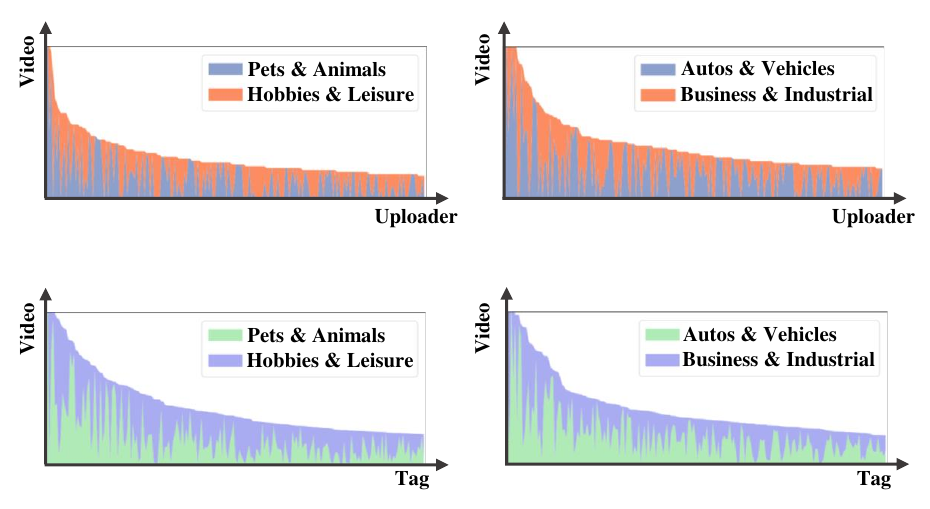}
\caption{Topic distribution for videos (1) from each uploader (Top) and (2) with which each tag has been associated (Bottom). $y$-axis is clipped to 80 / 1,000 for better illustration.}
\label{fig:stackplot}
\end{figure}

In the analysis, we first illustrate the distribution of tags for UGCs from two topics in both subsections of the dataset. From Fig. \ref{fig:comatrix} we can find that the two UGC topics in each sub-dataset are closely related as most tags are used in UGCs from both topics. This is desired to study the confounding bias \cite{gupta2021correcting}, because when topic $A$ and $B$ are highly related, uploaded videos may contain both $A$ and $B$ topics, where some uploaders may be more willing to assign $A$-topic-related tags while others might prefer $B$-topic-related tags. Under such cases, a deconfounded tag recommender can help them to select tags based solely on the UGC content irrespective of uploaders' personal bias. Then, we analyze uploaders and tags from the perspective of the distribution of (a) the topics of videos in each uploader's collections and (b) the topics of videos with which each tag has been associated, respectively. The two distributions are illustrated in Fig. \ref{fig:stackplot}. From the top part of Fig. \ref{fig:stackplot} we can determine that most uploaders tend to publish their UGC from both topic categories in both subsections of the YT-8M-Causal dataset. Since the uploaders induce the tag-selection bias which we would like to eliminate in DecTag, and an uploader's difference in tendency to select tags from one familiar category to describe her uploaded UGCs with cross-domain characteristics is one of the main sources of confounding bias in tag selections, this further demonstrates the value of the established YT-8M-Causal dataset to facilitate research that addresses confounding bias in tag recommendations. Similar analysis can be applied to tags illustrated at the bottom of Fig. \ref{fig:stackplot}.

\subsection{Causal Intervention on Test Sets}

After the preparation of all the data required by DecTag, our discussions shift to the intervention strategy applied to the test set such that the deconfounding ability of causal tag recommenders can be evaluated without the requirement for bias-free tags. 

\subsubsection{Intervening  on Test Set by Substitute Confounders}

As demonstrated in \cite{zheng2021disentangling}, the deconfounding ability of causal recommenders can be indirectly assessed by intervening in the distribution of treatment (UGC content in the case of tag recommendations) on the test set. The intervention is achieved by manually controlling the difference of treatments' distribution induced by confounders between the training set and test set. The rationale behind such an evaluation strategy is that, since the influence of uploaders is expected to be eliminated by an ideal deconfounded tag recommender system, the system selects tags based solely on the content of UGCs; therefore, recommendation accuracy should not degenerate on the intervened test set. Otherwise, the spurious correlation induced by uploaders is mistakenly captured by the system as the causal influence from the UGC content, where the recommendation performance will unavoidably decrease when the correlation changes in the test set. In addition, since in most cases, intervention cannot be achieved by directly changing the value of confounders, we intervene upon substitute confounders that have similar effects to the true confounders, \textit{i.e.,} the uploaders. Through this mechanism, the dataset intervention strategy avoids the intractable requirement of obtaining the bias-free tags for each uploaded UGC, which is a tractable means to evaluate causal tag recommender systems. 

Specifically, we choose the UGCs' topic as the substitute confounder upon which we conduct intervention because it simultaneously influences the content of UGC the uploaders tend to create and the tags selected for the content. Consequently, we assume that UGC topics could exert influences on tag selections similar to uploaders as true confounders, and controlling the UGC topic distribution as surrogate confounder provides us with a viable strategy to intervene the dataset according to the influence of true confounders. In our implementation, we split both subsections of the YT-8M-Causal dataset into training, validation, and test sets such that the ratio of two selected topics is $X : (10-X)$ for the training set and $(10-X) : X$ for the validation and test sets. $X \in \operatorname{range}(1, 9)$ represents the ratio of UGCs from the first topic category. In particular, $X$ controls the strength of intervention conducted upon the test set, where $X$ with a value close to extremity 1 or 9 demonstrates a stronger intervention imposed upon the test set.  

\subsubsection{Further Discussions} It is fully possible to select more than two UGC topics to establish the YT-8M-Causal dataset. We select only two categories of topic with intimate cross-topic relations because in such cases, more severe confounding bias exists in collected UGC-tag pairs, where deconfounded tag recommenders can be of great benefit as they increase UGCs' exposure to the right target audience interested in both topics by eliminating uploaders' personal bias in tag selections. In contrast, if more topics are included, some of the topics may not be related to each other, where deconfounding brings about fewer advantages in comparison. 

\section{Experiments}

\begin{table*}[t]
\caption{Comparisons between DecTag with other debias strategies on two base recommendation methods, NFM and LightGCN, on the established YT-8M-Causal datasets. Method with the highest performance is highlighted in \textbf{Bold}. We also conduct single-sided t-test on the results and report the p-values with base model to demonstrate the significance of the improvement.}
\label{tab:comparison}
\begin{center}
\setlength{\tabcolsep}{1mm}{
\resizebox{\textwidth}{!}{
\begin{tabular}{l|cccc|cccc|cccc|cccc}
\toprule
\textbf{} & \multicolumn{8}{c|}{\textbf{NFM}} & \multicolumn{8}{c}{\textbf{LightGCN}} \\ \hline 
\multirow{2}{*}{\textbf{Method}} & \multicolumn{4}{c|}{\textbf{YT-8M-Causal (P.A. \& H.L.)}} & \multicolumn{4}{c|}{\textbf{YT-8M-Causal (A.V. \& B.I.)}} & \multicolumn{4}{c|}{\textbf{YT-8M-Causal (P.A. \& H.L.)}} & \multicolumn{4}{c}{\textbf{YT-8M-Causal (A.V. \& B.I.)}} \\
 & \textbf{R@10} & \textbf{R@20} & \textbf{N@10} & \textbf{N@20} & \textbf{R@10} & \textbf{R@20} & \textbf{N@10} & \textbf{N@20} & \textbf{R@10} & \textbf{R@20} & \textbf{N@10} & \textbf{N@20} & \textbf{R@10} & \textbf{R@20} & \textbf{N@10} & \textbf{N@20} \\ \hline 
\textbf{None}         & 0.2189   & 0.3069   & 0.2002   & 0.2289   & 0.1900 & 0.2673 & 0.1680 & 0.1928    & 0.2320   & 0.3079   & 0.2244   & 0.2450   & 0.1920   & 0.2643   & 0.1786   & 0.2001 \\
\textbf{Unawareness}  & 0.2217   & 0.3125  & 0.2112   & 0.2327   & 0.1923  & 0.2713   & 0.1804 & 0.2023   & 0.2376   & 0.3124   & 0.2308   & 0.2509   & 0.2028   & 0.2812   & 0.1888   & 0.2135 \\
\textbf{FairCo}       & 0.2166   & 0.3055   & 0.1996   & 0.2286   & 0.1877   & 0.2623   & 0.1700    & 0.1935   & 0.2308   & 0.3127   & 0.2230   & 0.2459  & 0.1907   & 0.2629   & 0.1798   & 0.2017  \\
\textbf{Calibration}  & 0.2147   & 0.3054   & 0.1945   & 0.2192    & 0.1879   & 0.2662   & 0.1612    & 0.1856     & 0.2318   & 0.3073   & 0.2237   & 0.2442   & 0.1916   & 0.2634   & 0.1779   & 0.1992 \\ 
\textbf{IPS}          & 0.2210   & 0.3123   & 0.2050   & 0.2348   & 0.1919 & 0.2703 & 0.1717 & 0.1968    & 0.2385   & 0.3145   & 0.2283   & 0.2496   & 0.1997   & 0.2717   & 0.1843   & 0.2065 \\ 
\textbf{CR}    & 0.2226   & 0.3133   & 0.2078   & 0.2362   & 0.1931   & 0.2717   & 0.1745    & 0.2004   & 0.2414 & 0.3178 & 0.2331 & 0.2540    & 0.2046 & 0.2742 & 0.1939 & 0.2151  \\
\textbf{DecRS}        & 0.2255   & 0.3150   & 0.2098   & 0.2382   & 0.1960 & 0.2737 & 0.1802 & 0.2046    & 0.2434   & 0.3213   & 0.2377   & 0.2578   & 0.2125   & 0.2815   & 0.1988   & 0.2224   \\ \hline
\textbf{DecTag}       & \textbf{0.2314}   & \textbf{0.3200}   & \textbf{0.2233}   & \textbf{0.2505}   & \textbf{0.1974} & \textbf{0.2750} & \textbf{0.1838} & \textbf{0.2082}    & \textbf{0.2492}   & \textbf{0.3291}   & \textbf{0.2431}   & \textbf{0.2640}   & \textbf{0.2168}   & \textbf{0.2937}   & \textbf{0.2020}   & \textbf{0.2250}   \\
\textit{p-value}    & 5.71E-2   & 6.35E-2   & 1.03E-2   & 7.58E-3   & 1.11E-1    & 3.75E-2    & 1.12E-2     & 6.93E-3    & 2.12E-4   & 2.27E-4   & 9.15E-4   & 4.51E-4   & 8.64E-6   & 5.07E-5   & 4.66E-5   & 1.01E-4   \\
\bottomrule
\end{tabular}
}}
\end{center}

\end{table*}

In this section, we report the extensive experiments conducted on the established YT-8M-Causal dataset to demonstrate the effectiveness of the proposed DecTag. Specifically, we focus on investigating the three important research questions as follows:

\begin{itemize}[leftmargin=2em]
    \item \textbf{RQ1:} How does the proposed DecTag perform on various base models compared with other deconfounding strategies in terms of recommendation accuracy and robustness to confounding bias on the established YT-8M-Causal datasets?
    \item \textbf{RQ2:} How does the performance of different deconfounding strategies change with the increase of intervention strength (\textit{i.e.,} the difference of UGC topic distribution in the training and test sets) on the YT-8M-Causal test sets?
    \item \textbf{RQ3:} Could a large batch size indeed help DecTag reduce the estimation variance of the proposed MC-based do-calculus estimator with bootstrap when sample size $N^{s}$ is small?
\end{itemize}

\subsection{Experimental Setups}

\subsubsection{Training Strategy}

As we have discussed in Section \ref{sec:trade-off}, a large batch size is equivalent to a large sample size for the MC-based do-calculus to reduce the estimation variance. Therefore, batch size is set to 1,024 for NFM backbone and 4,096 for LightGCN backbone to stabilize the training dynamics, which is larger than batch size normally used in tag recommendations. $N^{s}$ is set to 1. We use Adam optimizer \cite{kingma2015adam} with learning rate 1e-3. Five epochs are used to learn the conditional distribution to warm up the model (Section \ref{sec:conf_form}), and the deconfounded training stops at the 200\textit{th} and 400\textit{th} epoch for NFM and LightGCN backbone, respectively. The parameter that controls dataset intervention strength, $X$, is set to 3 for comparisons with other deconfounding/debiasing strategies in Section \ref{sec:comp}, and its effects are thoroughly discussed in Section \ref{sec:interv}.

\subsubsection{Evaluation Metrics}

In this paper, the tag recommendation accuracy is measured by two commonly-used ranking metrics, \textit{i.e.,} Recall@K (R@K) and NDCG@K (N@K). As we have demonstrated in Section \ref{sec:trade-off}, a higher accuracy on the intervened test set where the UGC topic distribution differs from that of the training set can also demonstrate the deconfounding ability of different strategies under certain assumptions. Therefore, these two metrics can also be used to compare the deconfounding ability of different strategies.

\begin{figure*}
\setlength{\abovecaptionskip}{0.15cm}
\setlength{\belowcaptionskip}{0.1cm}
\centering
\includegraphics[width=\linewidth]{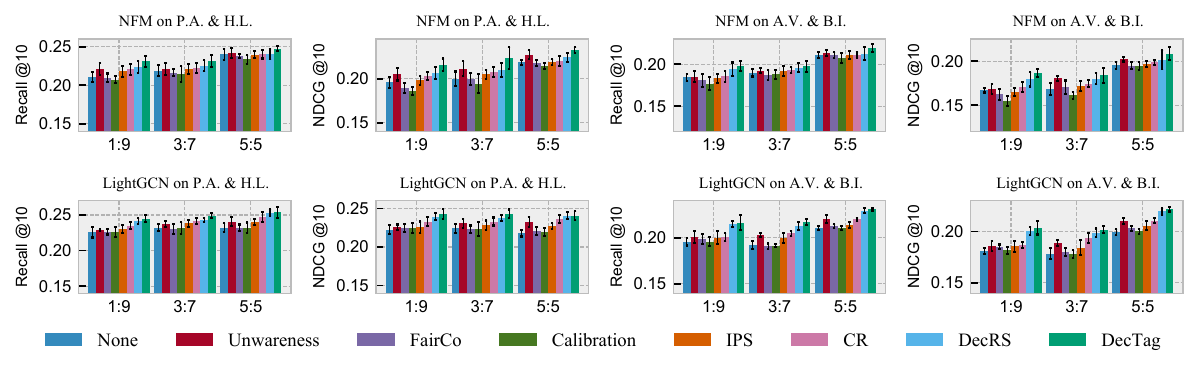}
\caption{Comparisons between DecTag and different deconfounding/debiasing strategies on the established YT-8M-Causal dataset with varied test set intervention strength. Error bar shows the standard deviation on five splits of the dataset.}
\label{fig:breakdown}
\end{figure*}

\subsection{Comparisons with Baselines}

\subsubsection{Baseline Descriptions}

To answer \textbf{RQ1}, we compare the proposed DecTag with state-of-the-art debiasing and deconfounding strategies on two widely-adopted baseline recommendation models, \textit{i.e.,} NFM \cite{he2017neural} and LightGCN \cite{he2020lightgcn}, as follows:

\begin{itemize}[leftmargin=2em]
    \item \textbf{None}. This is the naive strategy where no debiasing strategy is applied to the baseline models. Specifically, the strategy directly fits the conditional distribution $P(T \mid U, C)$ on the uploader-confounded dataset to predict tags for new UGCs.
    \item \textbf{Unawareness} \cite{grgic2016unawareness}. A tag recommender system with Unawareness is achieved by explicitly excluding the user-related features from the model. Therefore, it fits the conditional distribution $P(T \mid C)$ on the confounded dataset where only content features of UGCs are considered for tag predictions.
    \item \textbf{FairCo} \cite{morik2020fairco}.  This is the dynamic Learning-to-Rank algorithm which introduces an error term to control exposure-based fairness among different groups. We divide tags into different category groups based on historical data and calculate the error term based on the ranking list sorted by relevance.
    \item \textbf{Calibration} \cite{steck2018calibrated}. This method alleviates bias amplification by introducing and controlling a calibration metric $C_{KL}$ that measures the imbalance between interaction history and recommendation list. We calculate $C_{KL}$ based on the historical tags of uploaders and minimize it by re-ranking. 
    \item \textbf{IPS} \cite{saito2020ips}.  IPS aims to eliminate confounding bias by reweighting the samples by the chance of their observation (\textit{i.e.,} propensity score). The propensity score of each UGC is estimated from its uploader's representation. In addition, a clipping technique \cite{saito2020ips} is introduced to reduce the estimation variance.
    \item \textbf{CR} \cite{wang2021clicks}. CR was originally designed to address the clickbait issue by conducting counterfactual inference on causal graphs. In this work, similar procedures can be applied to estimate the matching likelihood of a UGC-tag pair in a counterfactual world such that the effect of uploaders' bias can be reduced.
    \item \textbf{DecRS} \cite{wang2021deconfounded}. DecRS is a deconfounded recommender system where deconfounding (in the case of tag recommendation) is achieved by substituting a hypothetical averaged uploader for the true uploader when training the model, such that the bias of a specific uploader can be reduced by averaging.
\end{itemize}

For a fair comparison, the base model structures and the hyperparameters of both base models and deconfounding/debiasing strategies, such as the embedding size, the batch size, $\lambda$ in the ranking target for FairCo and Calibration, and the clipping threshold $t$ for IPS, are systematically determined through grid research. 

\subsubsection{Comparison Results and Analysis}
\label{sec:comp}
The comparison results are summarized in Table \ref{tab:comparison}.  Table \ref{tab:comparison} first reveals that the Unawareness strategy can improve over None where no deconfounding/debiasing is applied to the base model. Recall that both strategies fit biased models (\textit{i.e.,} conditional distributions) on the confounded dataset with the discrepancy that Unawareness ignores the uploader and uses only UGC content as the condition. This may suggest that uploaders indeed exert heavy influence on tag selections, where uploader-induced confounding bias is exacerbated. However, Unawareness can only address explicit uploader-related confounding bias, because UGC content might be further disentangled into useful information and implicit confounders such that spurious association endures even with the removal of uploaders. Consider again the toy examples introduced in Section \ref{sec:causal_graph}. Suppose that video uploaders who bias to nature-related tags also tend to include green colors (from forests, prairie, or meadows, \textit{etc.}) in their videos. Under such cases, even if Unawareness ignores uploaders in its conditional modeling, part of the UGC content, \textit{i.e.,} the green color in the example, can still serve as a surrogate for uploaders that spuriously correlates UGC content and tags. Moreover, we also find that Calibration and FairCo may perform worse than the backbone models, which suggests that simple re-ranking or group fairness constraint without causality may hurt tag recommendation models' performance and robustness to confounding bias \cite{ziegler2005improving, wang2021deconfounded}. 

With causal inference techniques explicitly introduced to address uploaders' confounding bias, the performance of two base models improves for CR and DecRS compared to Unawareness and Calibration. Although causality has been employed in IPS, which deconfounds by reweighting the UGC-tag pairs based on the inverse of propensity scores estimated from uploader representations, it achieves the worst performance among all causality-based methods. The reason could be that the unbiasedness of IPS requires a correctly specified score estimation model, which relies heavily on the expertise of the researchers. Moreover, even with clipping strategies, IPS may lead to unstable training dynamics if the estimated propensity scores vary drastically for different samples. Conducting counterfactual inference on the causal graph to estimate, CR improves over IPS, due to the comparatively low variance of counterfactual reasoning to estimate the likelihood of UGC-tag pairs. Utilizing a hypothetical averaged uploader to replace the true biased uploader for backdoor adjustment, DecRS performs the best among all included deconfounding strategies. However, the unbiasedness of DecRS requires two strong assumptions, \textit{i.e.,} no residual bias and a linear relation between tag selections and uploader representations, both of which, as we have analyzed in Section \ref{sec:trade-off}, may not hold in the case of tag recommendations.

The proposed DecTag achieves better performance than DecRS on both datasets and improves significantly over the base models. This demonstrates the superiority of MC-based do-calculus approximator with bootstrap to solve the backdoor adjustment on causal graphs for tag recommendations, where unbiasedness can be achieved with weaker assumptions compared with DecRS, \textit{i.e.,} the uploaders are \textit{i.i.d.} samples from the population. The training time complexity, when using a larger batch size instead of a larger sample size for the MC estimator, increases only slightly compared with DecRS and {can be effectively vectorized by a single GTX1080 GPU with 8G memory}. In addition, since the MC-based approximator can be viewed as replacing the biased uploader with an unbiased hypothetical random uploader when making recommendations, the deconfounding of uploaders' influences in DecTag can be achieved in an interpretable manner. Therefore, these clearly demonstrate the advantage of DecTag compared with other state-of-the-art deconfounding/debiasing strategies for UGC tag recommendations.

\subsection{Sensitivity to Dataset Intervention}
\label{sec:interv}
To answer \textbf{RQ2}, we vary the strength of intervention imposed on the YT-8M-Causal test sets, \textit{i.e.,} the ratio of the two topics that we intervene upon and evaluate different deconfounding/debiasing strategies. Specifically, we vary $X \in \{\operatorname{range}(1,5,2)\}$ and conducted similar experiments as Section \ref{sec:comp}. The comparison results are illustrated in Fig. \ref{fig:breakdown}. From Fig. \ref{fig:breakdown} we can find that with the increase of intervention strength conducted upon the test set, the performance decreases for all deconfounding/debiasing strategies for both the NFM and LightGCN backbones. This could either indicate that there exists a systematic difference between the accuracy of tag recommendation for UGCs from the two topics in each subsection of the dataset, or residual confounding bias exists that fails to be eliminated by the methods. However, the best performance of the proposed DecTag achieved with all intervention strength more convincingly demonstrates the superior deconfounding ability of the proposed DecTag in comparison with the state-of-the-art deconfounding/debiasing strategies for tag recommendations.

\subsection{Sensitivity to Sample Size of MC-Estimator}

\begin{figure*}
\setlength{\abovecaptionskip}{0.1cm}
\setlength{\belowcaptionskip}{-0.1cm}
\centering
\includegraphics[width=0.97\linewidth]{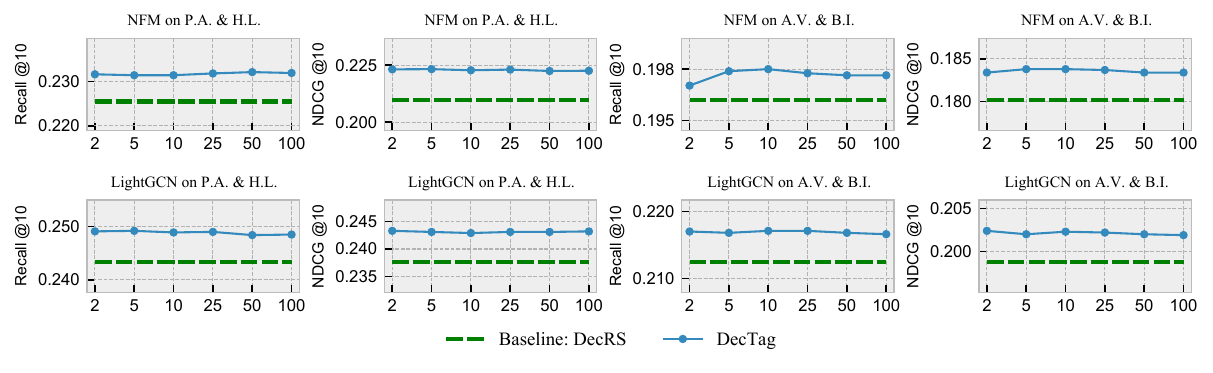}
\caption{The sensitivity of DecTag's performance to the sample size $N^{s}$ of the proposed MC-based do-calculus approximator when the batch size is large (\textit{i.e.,} 1,024 for NFM base model and 4,096 for LightGCN base model).}
\label{fig:sample_size}
\end{figure*}

In \textbf{RQ3}, the sample size $N^{s}$ for the MC-based do-calculus approximator is a very important hyperparameter that controls the trade-off between estimator variance and computational overhead. In Section \ref{sec:trade-off}, we claim that a large batch size can reduce the variance similar to a large $N^{s}$, based on previous literature \cite{kingma2013auto}. In this section, we conduct experiments to verify such a claim. Specifically, we keep the large batch size for both base models (\textit{i.e.,}  1,024 for NFM and 4,096 for LightGCN), vary $N_{s} \in \{2, 5, 10, 25, 50, 100\}$, and report the performance in Fig. \ref{fig:sample_size}. From Fig. \ref{fig:sample_size} we can find that the performance is indeed stable with varied $N^{s}$. Such results indicate that a large batch size can reduce the variance and stabilize training equivalent to a large sample size. In addition, setting a large batch size increases only slightly the training complexity compared with DecRS, which more convincingly shows the superiority of DecTag.

\section{Related Work}

Although DecTag is the first causality-based algorithm for content-based UGC tag recommendation to our knowledge, it draws inspiration from previous tag recommender systems and current trends in causal item recommendations, which we survey as follows.

\vspace{0.2cm}  
\noindent \textbf{Tag Recommender Systems.} UGC tag recommendation aims to automatically recommend tags to users when they upload generated content into online platforms \cite{song2011automatic,belem2017survey}. Existing tag recommenders can be classified into collaborative-filtering (CF)-based methods \cite{marinho2008collaborative} and content-based methods \cite{lu2009content}. CF-based methods \cite{marinho2008collaborative} recommend new tags to a UGC with pre-existing tags based on UGCs with similar tags. Since collaborative information is scarce, content-based tag recommendation is more common where tags are suggested by UGC content \cite{lu2009content}. However, both strategies merely learn the association between existing UGC-tag pairs. Since the uploader simultaneously creates the UGC content and selects its tags, she is a confounder that creates a spurious correlation between UGC and the self-selected tags. This may lead to systematic bias in traditional tag recommender systems, which can be addressed by the causal inference techniques introduced in this paper. 

It can be noticed that another trend in tag recommender system research is towards personalization \cite{rendle2010pairwise, guan2009personalized,zhao2021personalized}, where uploaders' preferences are explicitly utilized to recommend tags for their generated content. Although this may elicit familiar feelings from the uploaders due to the explicit consideration of their personal preferences and habits, the recommended tags may end up catering to the uploaders' personal bias instead of finding the most suitable tags for the content of the UGC, which may exacerbate the bias and cause uploaders to lose exposures to certain audiences.

\vspace{0.2cm}
\noindent \textbf{Recommendations with Causality.} Pearl's causal model is widely used for item recommendations \cite{pearl2016causal}. In Pearl's framework, relations among variables are represented by graphs, where backdoor adjustment is conducted on the treatment node (usually item exposures) such that spurious correlation induced by confounders can be eliminated. Confounders usually include item attributes \cite{wang2020causal, feng2021should}, user preferences \cite{wang2021deconfounded} and historical interactions \cite{bonner2018causal,xu2021causal}. Existing solutions to backdoor adjustment can be categorized into inverse propensity-score reweighting (IPS)-based methods \cite{schnabel2016recommendations,zou2020counterfactual,lee2021dual} and approximation-based methods \cite{wang2021deconfounded}. IPS reweights samples by their observation probabilities such that they can be viewed as uniformly selected from the population, irrespective of confounders' influences. However, the unbiasedness of IPS requires a correctly specified score estimation model. Moreover, it leads to unstable training dynamics if scores vary drastically among samples. In contrast, approximation-based methods substantially reduce the training variance, but most existing strategies will introduce extra bias if non-linear relations exist between confounders and ratings  \cite{wang2021deconfounded}. 

In addition, compared with item recommendations, causal inference in tag recommendation tasks faces new challenges, such as complex treatment distributions (where treatments are continuous UGC content instead of binary item exposures), complex confounders (where confounders are UGC uploaders instead of categorical item attributes or user preferences), and complex relations among treatments, confounders, and outcomes (\textit{i.e.,} UGC content, uploader, and tags). Therefore, the above challenges clearly motivate us to design DecTag, where the deconfounding of uploaders' influence can be achieved with superior performance, asymptotic consistency and unbiasedness, and good interpretability.

\section{Conclusions}

In this paper, we have proposed a deconfounded tag recommender system, \textit{i.e.,} DecTag, based on causal intervention. Specifically, we derive a Monte Carlo-based backdoor adjustment strategy, where uploader-induced confounding bias can be interpretably addressed by substituting a hypothetical random uploader for the true uploader with personal bias. In addition, to evaluate the deconfounding ability of different strategies, a large-scale tag recommendation dataset, \textit{i.e.,} YT-8M-Causal, is established where an evaluation strategy is introduced accordingly. Extensive experiments demonstrate that DecTag is more robust to uploaders' confounding bias than state-of-the-art deconfounded tag recommendation strategies.

\balance

\bibliographystyle{ACM-Reference-Format}
\bibliography{dectag}

\end{document}